
\hfuzz=1pt
\input phyzzx
\PHYSREV
\FRONTPAGE

\def\pr#1#2#3{Phys. Rev. D{\bf #1}, #2 (19#3)}
\def\prl#1#2#3{Phys. Rev. Lett. {\bf #1}, #2 (19#3)}

\def\np#1#2#3{Nucl. Phys. {\bf B#1}, #2 (19#3)}
\def\pl#1#2#3{Phys. Lett. {\bf #1B}, #2 (19#3)}


\hfill {CU-TP-564}\break
\strut\hfill {May 1992}
\vglue 1in
\font\twelvebf=cmbx12
\vskip 1.0in
\centerline {\twelvebf Self-Dual Chern-Simons Higgs Systems}
\centerline {\twelvebf with an  N=3 Extended  Supersymmetry
\footnote{}{\rm This work
was supported in part by the U.S. Department of Energy (HK), an NSF
Presidential Young Investigator Award (KL) and an Alfred P. Sloan
Fellowship (KL).} }
\vskip .5in
\centerline{\it  Hsien-Chung  Kao and
 Kimyeong Lee }
\vskip .1in
\centerline {Physics Department, Columbia University}
\centerline {New York, N.Y. 10027}
\vskip .4in
\centerline{\bf Abstract}

We construct and study an $N=3$ supersymmetric Chern-Simons Higgs
theory. This theory is the maximally supersymmetric one containing the
self-dual models with a single gauge field and no gravity.

\endpage

\REF\rJW{ J. Hong, Y. Kim, and P.Y. Pac, \prl{64}{2230}{90}; R. Jackiw
and E.J. Weinberg, \prl{64}{2234}{90}; R. Jackiw, K. Lee, and E.J.
Weinberg, \pr{42}{3488}{90}. }
\REF\rLLW{ C. Lee, K. Lee and E.J. Weinberg, \pl{243}{105}{90}; P.P.
Srivastava and K. Tanaka, \pl{256}{427}{91}; E.A.
Ivanov, \pl{268}{203}{91}. }
\REF\rWitten{ E. Witten and D. Olive, \pl{78}{97}{78}; P. Di Veccia and
S. Ferrara, \np{130}{93}{77}.}
\REF\rSUP{G.V. Grigoryev and D.I. Kazakov, \pl{253}{411}{91};
 L.V. Avdeev, G.V. Grigoryev, D.I. Kazakov, CERN preprint,
CERN-6091/91 (1991);  Y. Ipekoglu, M. Leblanc, and M.T. Thomaz,
MIT preprint, CTP \#1987 (1991). }

\REF\rFIN{ S. Mandelstam, \np{213}{149}{83}; P.S. Howe, K.S. Stelle and
P. West, \pl{124}{55}{83}.}
\REF\rWest{ P. West, {\it Introduction to Supersymmetry and
Supergravity},  the 2 edition (World Scientific, 1990). }

\REF\rGates{S.J. Gates, M.T. Grisaru, M. Ro{\v c}ek and W. Siegel,
{\it Superspace, or One Thousand and One Lessons in Supersymmetry},
(Benjamin/Cummings, Reading, MA, 1983). }

\REF\rWess{ J. Wess and J. Bagger, {\it Supersymmetry and Supergravity},
(Princeton Univ. Press, Princeton, NJ, 1983). }

\REF\rQED{ J. Wess and B. Zumino, \np{78}{1}{74}.}

\REF\rColeman{ S. Coleman and B. Hill, \pl{159}{184}{85}. }

\chapter{ Introduction }

Recently there has been considerable interest in self-dual
Chern-Simons-Higgs systems in 2+1 dimensional spacetime. With a
specific sixth-order Higgs potential which has degenerate symmetric
and asymmetric minima, it has been shown that there are topological
vortices and nontopological solitons satisfying self-dual or
Bogomolny-type equations.\refmark{\rJW} This Higgs potential is shown
to be determined by requiring an $N=2$ extended
supersymmetry.\refmark{\rLLW} The central charge of the extended
supersymmetric algebra gives a quantum version of the Bogomolny
bound.\refmark{\rWitten} Renormalization group and finite temperature
properties of this $N=2$ supersymmetric theory has been
studied.\refmark{\rSUP }

It is well known that a theory with extended supersymmetry has less
ultraviolet divergence. Some supersymmetric theories have  been shown
to be even finite.\refmark{\rFIN, \rWest} It would be interesting to know
whether there is a supersymmetric Chern-Simons Higgs theory which is
ultraviolet finite.  It is not so hard to see that the $N=2$
supersymmetric theory in Ref.[2] is not finite perturbatively.
Motivated partially by this observation,  we construct in this paper
an $N=3$ supersymmetric Chern-Simons Higgs theory and study some
properties of it.  This theory turns out to be the `maximally
supersymmetric' Chern-Simons Higgs theory with a single
gauge field and no gravity.

In four dimensional spacetime, an $N=3$ supersymmetric theory becomes
$N=4$ supersymmetric because of the $CPT$ and rotational symmetry.
(For a general reference for supersymmetry, see for example Ref.[6,7,8].)
In three dimensional spacetime a particle and its antiparticle
carry the same spin, say $s$. Under rotation a state of angular
 momentum $j$  transforms into itself with some phase. There is no
need to have a state of spin $-s$ in the theory.
If we represent anyons as charged particles in a Chern-Simons Higgs
theory, both anyons and antianyons have spin $s=1/(4\pi \kappa)$ where
$\kappa$ is the coefficient of the Chern-Simons term. They can
annihilate each other into neutral integer spin particles because the
orbital angular momentum between them is $-2s + integer$. Hence in
three dimensional spacetime there can be an $N=3$ supersymmetric theory.

In section 2 we construct an $N=2$ supersymmetric Chern-Simons Higgs
theory in three dimensional spacetime by applying the dimensional
reduction method to the four dimensional $N=1$ supersymmetric QED.
This $N=2$ theory does not have any central charge and is different
from the theory in Ref.[2].  In section 3 we eliminate the auxiliary
fields from the lagrangian and choose a special set of coupling
constants, resulting in an additional global $U(1)$ symmetry.  We use
this new global symmetry to complexify one of $N=2$ supersymmetries,
leading to an $N=3$ extended supersymmetry.  This is quite similar to
the method used in Ref.[2] to get the $N=2$ supersymmetry.  In section
4 we study some properties of the theory.  In section 5 we investigate
the $N=3$ supersymmetric operator algebra.  There is a quantum version
of the Bogomolny-type bound on the energy functional. We show that the
maximally supersymmetric Chern-Simons Higgs theory with a single gauge
field and no gravity is $N=3$.  In section 6, we conclude with some
remarks.

\endpage

\chapter{ N = 2 Supersymmetry }

By the dimensional reduction method, one can obtain a theory of larger
supersymmetry.  For example, the $N=1$ supersymmetric $QED$ in four
dimensional spacetime will lead to a $N=2$ supersymmetric $QED$ in
three dimensional spacetime.  The dimensional reduction method
specifies both the lagrangian and the supersymmetric transformation
law.  We start by reviewing the $N=1$ supersymmetric $QED$ in four
dimensional spacetime.\refmark{\rQED} The gauge multiplet is made of $A_\mu,
\lambda, D$ and the chiral matter multiplet is made of $z_k,
\psi_k, f_k,$ where $k=1,2$ denote the components of a vector under
the $U(1)$ gauge symmetry. The sign of the metric is chosen to be
$(-,+,+,+)$.  The gamma matrices  in the Majorana
representation are
$$ \gamma^0 = -i \left( \matrix{ 0 & \sigma^2 \cr
                             \sigma^2 & 0 \cr} \right),\,\,
 \gamma^1 = \left( \matrix{ 0 & \sigma^3 \cr
                           \sigma^3 & 0 \cr} \right),\,\,\,
  \gamma^2 = \left( \matrix{ 0 & \sigma^1 \cr
                             \sigma^1 & 0 \cr} \right), \,\,\,
  \gamma^3 = \left( \matrix{ {\bf 1} & 0 \cr
                              0 & -{\bf 1} \cr} \right). $$
We use the convention where $\gamma_5 = \gamma_0 \gamma_1 \gamma_2
\gamma_3$ and $\bar{\psi} = \psi^\dagger \gamma^0$.
The fermion fields are Majorana or $\lambda^* =
\lambda, \psi_k^* = \psi_k$.

The covariant derivative of the matter multiplet will be given by $
D_\mu z_k = \partial_\mu z_k - e A_\mu z^k $, where $z^1= z_2$ and
$z^2 = -z_1$.  The standard $N=1$ supersymmetric kinetic term for a
charged matter multiplet coupled to the gauge multiplet is then
$$ {\cal L}_{K} = -|D_\mu z_k|^2 + |f_k|^2 -
i\bar{\psi}_k\gamma^\mu D_\mu \psi_k
 + i eD  z_k z^{k*}  +  2i e \bar{\lambda}[ \psi_{kL}z^{k*} + \psi_{kR}
z^k ] \eqno\eq $$
where $\psi_{L,R}  = (1 \pm  i\gamma_5)\psi /2 $.  Besides the $U(1)$
gauge symmetry, the kinetic term has an additional chiral $U(1)$
symmetry, which changes the phases of $z_k, f_k^*, \psi_{kL},
\psi_{kR}^*$ uniformly.

The self-interaction of the matter multiplet is given by the gauge
invariant superpotential $W$, which  as a function
of $z_k$ is given by
$$ W =  { m \over 2} (z_k)^2 + {g\over 4} (z_k^2)^2 \eqno\eq $$
where the coupling constants $m, g$ are complex. From the above
superpotential  one can obtain the supersymmetric self-interaction
lagrangian for the matter multiplet,
$$ {\cal L}_P = {\delta W \over \delta z_k} f_k - i { \delta^2 W \over
\delta z_k \delta z_l} \bar{\psi}_{kL}^* \psi_{lL} + h.c. \eqno\eq $$
This self-interaction becomes renormalizable only after dimensional
reduction to three dimensional spacetime.

The kinetic and interacting terms of the matter multiplet in Eqs. (2.1)
and (2.3) are invariant under the $N=1$ supersymmetric transformation.
The gauge multiplet transforms as
$$ \eqalign{ &\ \delta A_\mu = i \bar{\alpha} \gamma_\mu \lambda \cr
&\  \delta \lambda = -{1\over 2} F_{\mu\nu} \gamma^{\mu\nu} \alpha +
D \gamma_5 \alpha  \cr
&\  \delta D = i \bar{\alpha} \gamma_5 \gamma^\mu \partial_\mu \lambda
\cr} \eqno\eq $$
where the parameter $\alpha$ is a Majorana spinor.
The matter multiplet transforms as follow,
$$ \eqalign{ &\  \delta z_k = 2 i \bar{\alpha} \psi_{kL} \cr
&\  \delta \psi_{kL} = D_\mu z_k \gamma^\mu \alpha_R + f_k \alpha_L \cr
&\  \delta f_k = 2i \bar{\alpha}\gamma^\mu D_\mu \psi_{kL}
        -2ie \bar{\alpha} \lambda_R z^k  \cr} \eqno\eq $$

Dimensional reduction to three dimensional spacetime is done by
assuming that the fields are independent of the third spatial
coordinate. In three dimensional spacetime, we keep the same notation
for the gamma matrices. This will not cause any confusion because all
the following derivations and calculation will be in three dimensional
spacetime.  The gamma matrices in the three dimensional spacetime are
again in the Majorana representation and are given by
$$ \gamma^0 = -i\sigma^2, \gamma^1 = \sigma^3, \gamma^2 = \sigma^1, $$
which satisfy $\gamma^\mu \gamma^\nu = \eta^{\mu\nu} +
\epsilon^{\mu\nu\rho}\gamma_\rho $ with $\epsilon^{012} = 1$.
We then rename the third component of the gauge field and split
four-component spinors into two two-component spinors in three dimension;
$$  A_3 = C, \,\,\, \lambda = { \lambda_1 \choose \lambda_2 },
\,\,\, \alpha = {\alpha_1 \choose  \alpha_2} $$

Let us first apply the dimensional reduction method to the
supersymmetric transformation of the gauge multiplet.  The
supersymmetric transformation (2.4) becomes
$$\eqalign{&\  \delta A_\mu = i \bar{\alpha}_a \gamma_\mu \lambda_a \cr
&\  \delta \lambda_a = -B^\mu \gamma_\mu
 \alpha_a  + \partial_\mu C \gamma^\mu \alpha^a + D\alpha^a \cr
&\  \delta C = i  \bar{\alpha}^a \lambda_a \cr
&\  \delta D = i\bar{\alpha}^a \gamma^\mu \partial_\mu \lambda_a
\cr}\eqno\eq $$
where $B^\mu = \epsilon^{\mu\nu\rho}F_{\nu\rho}/2$, $a=1,2$, and
$\alpha^1 = \alpha_2, \alpha^2 = - \alpha_1 $.  With two independent
parameters $\alpha_1, \alpha_2$, the supersymmetric transformation
becomes $N=2$ in three dimensional spacetime.  $A_\mu, \lambda_a, C,D$
form an $N=2$ supersymmetric vector multiplet.  We are interested in
Chern-Simons rather than Maxwell kinetic term.  The $N=2$
supersymmetric version of the Chern-Simons kinetic term can be easily
guessed by using the supersymmetric transformation (2.6) and is
 $$ {\cal L}_{CS} = {\kappa \over 2}
(\epsilon^{\mu\nu\rho} A_\mu
\partial_\nu A_\rho - i\bar{\lambda}_a\lambda_a + 2CD). \eqno\eq $$

The demensional reduction of the scalar multiplet is facilitated by
redefining fields,
 $$ \eqalign{&\ \phi_1 = {\rm Re}\, z_1 + i
{\rm Re}\, z_2 \cr &\ \phi_2 = {\rm Im}\, z_1 + i {\rm Im}\, z_2 \cr
&\ F_1 = {\rm Re}f_1 + i{\rm Re}f_2 \cr &\ F_2 = -{\rm Im}f_1 - i {\rm
Im}f_2 \cr &\ {\psi \choose \chi} = \psi_1 + i \psi_2 \cr } \eqno\eq$$
where $\psi, \chi$ are two-component spinors. The complex fields
$\phi_a, \psi,\chi, F_a$, form an $N=2$ supermultiplet in three
dimensional spacetime and carry unit electric charge. The
transformation (2.5) becomes  the $N=2$ supersymmetric transformation,
$$\eqalign{ &\ \delta \phi_a = i(\bar{\alpha}^a\psi+
\bar{\alpha}_a \chi ) \cr
&\ \delta \psi = D_\mu \phi_a \gamma^\mu \alpha^a + F_a\alpha_a +
    ieC\phi_a \alpha_a \cr
&\  \delta \chi = D_\mu \phi_a \gamma^\mu \alpha_a + F_a\alpha^a -ie
C\phi_a \alpha^a \cr
&\  \delta F_a = i\bar{\alpha}_a\gamma^\mu D_\mu \psi + i
\bar{\alpha}^a \gamma^\mu D_\mu \chi  \cr
&\ \,\,\,\,\,\,\,\,\,\, - eC(\bar{\alpha^a}\psi -
\bar{\alpha}_a \chi) - e(\bar{\alpha}^a\lambda_b \phi_b +
\bar{\alpha}_a \lambda^b \phi_b) \cr} \eqno\eq $$
where $D_\mu \phi_a = \partial_\mu \phi_a + ieA_\mu \phi_a $, etc.
The kinetic term (2.1) becomes
$$\eqalign{  {\cal L }_K = &\  - |D_\mu \phi_a|^2 + |F_a|^2
 -i \bar{\psi}\gamma^\mu D_\mu \psi -i \bar{\chi}\gamma^\mu D_\mu
\chi  \cr
&\  -e^2 C^2 |\phi_a|^2  + eC(\bar{\chi}\psi - \bar{\psi}\chi)
+ieD \phi_a \phi^{a*} \cr
&\ + e\bar{\lambda}_a [ \psi \phi^{a*}  - \psi^*\phi^a - \chi \phi_a^*
+  \chi^* \phi_a ] \cr} \eqno\eq $$
The dimensional reduction of the  potential energy (2.3)
leads to
$$ \eqalign{  {\cal L}_P = &\ {1\over 2}
[m +  g (\phi_1 + i \phi_2)(\phi_1^* + i
\phi_2^* ) ][ (\phi_a+i \phi^a) F_a^* + (\phi_a^* +i\phi^{a*})F_a ]
 \cr
&\ +{i \over 2}[ m + 2g(\phi_1+ i\phi_2)(\phi_1^* +i \phi_2^*)]
[ i\bar{\psi}\psi  -i\bar{\chi}\chi - \bar{\chi}\psi -\bar{\psi}\chi]
\cr
&\   +{ig\over 4} (\phi_1 +i\phi_2)^2( i\bar{\psi}\psi^* -
i\bar{\chi}\chi^* - 2\bar{\chi}\psi^*  )  \cr
&\
+{ig\over 4}(\phi_1^* +i\phi_2^*)^2(i\bar{\psi}^*\psi -
i\bar{\chi}^*\chi - 2\bar{\chi}^*\psi ) + h.c. \cr} \eqno\eq $$

The $N=2$ supersymmetric lagrangian in three dimensional spacetime
is then the sum of the kinetic terms  in Eqs. (2.7) and (2.10) and
the potential term (2.11), which is given symbolically  by
$$ {\cal L}_{N=2} = {\cal L}_{CS} + {\cal L}_K + {\cal L}_P \eqno\eq $$
Note that ${\cal L}_{CS} + {\cal L}_K$ are invariant under the global
$O(2)$ rotation in $a$-indices.  In contrast to theory in Ref. [2], this
$R$-symmetry  is broken by ${\cal
L}_P$.

\endpage

\chapter{  N=3 Supersymmetry }

We now have an $N=2$ supersymmetric Chern-Simons-Higgs theory (2.12),
which has a $U(1)$ gauge symmetry.  Let us eliminate the auxiliary
fields $\lambda, C, D, F,$ from the $N=2$ supersymmetric lagrangian
(2.11) by using their field  equations,
$$ \eqalign{ &\ \lambda_a = -{ie \over \kappa} (\phi^{a*}\psi
 -\phi^a\psi^* -\phi_a^*\chi + \phi_a\chi^* ) \cr
&\ C= -{ie \over \kappa} \phi_a \phi^{a*} \cr
&\ D = -{2ie^3 \over \kappa^2}|\phi_a|^2 \phi_a \phi^{a*}
           -{e\over \kappa}(\bar{\chi}\psi -\bar{\psi}\chi) \cr
&\ F_a = -\phi_a[ m_R + g_R(|\phi_1|^2 -|\phi_2|^2)-
g_I(\phi_1\phi_2^*+\phi_1^*\phi_2) ] \cr
&\ \,\,\,\,\,\,\,\,\,\,\,\, +\phi^a[m_I+
g_I(|\phi_1|^2-|\phi_2|^2) + g_R(\phi_1\phi_2^*+ \phi_1^*\phi_2)] \cr}
\eqno\eq$$
where the subscripts $R, I$ to coupling constants denotes the real and
imaginary parts.

The terms depending on the gaugino field $\lambda_a$ become
$$\eqalign{{\cal L}_{\lambda} = {ie^2\over 2\kappa}[ &\ 2|\phi_a|^2
( \bar{\psi}\psi +\bar{\chi}\chi ) -
(\phi_a)^2( \bar{\psi}\psi^* +\bar{\chi}\chi^* ) \cr
&\ -(\phi_a^*)^2(\bar{\psi}^*\psi + \bar{\chi}^*\chi) -
2\phi_a\phi^{a*}(\bar{\chi}\psi -\bar{\psi}\chi)] \cr} \eqno\eq $$
The terms depending on $C,D,$   become
$$ {\cal L}_{CD} = {e^4 \over \kappa^2}|\phi_a|^2(\phi_a\phi^{a*})^2
 -{ie^2\over \kappa}\phi_a \phi^{a*}(\bar{\chi}\psi-\bar{\psi}\chi)
\eqno\eq $$
The terms depending on $F_a$ become
$$\eqalign{ {\cal L}_F =  &\ -|\phi_a|^2 \biggl\{  |m|^2 +
|g|^2(|\phi_1|^2-|\phi_2|^2)^2
+|g|^2( \phi_1\phi_2^*+\phi_1^*\phi_2)^2 \biggr.  \cr
&\  \biggl.  +(mg^*+ m^*g)(|\phi_1|^2 -|\phi_2|^2)
-i(mg^* -m^* g)(\phi_1 \phi_2^*+ \phi_1^* \phi_2) \biggr\}  \cr }\eqno\eq $$

 From Eqs. (2.7) and (2.10), we obtain the kinetic part of the
lagrangian,
$$ {\cal K } = {\kappa \over 2} \epsilon^{\mu\nu\rho}A_\mu
\partial_\nu A_\rho + |D_\mu \phi_a|^2 - i \bar{\psi}\gamma^\mu D_\mu
\psi - i\bar{\chi}\gamma^\mu D_\mu \chi \eqno\eq $$
We rewrite  the fermion-boson interacting part of Eq. (2.11),
$$ \eqalign{ {\cal L}_{fi} = &\ - \{ m_I + 2 g_I(|\phi_1|^2-|\phi_2|^2) +
2g_R(\phi_1 \phi_2^* + \phi_1^*\phi_2)\} (i\bar{\psi}\psi- i\bar{\chi}\chi
) \cr
&\ - \{ m_R + 2g_R(|\phi_1|^2 -|\phi_2|^2) -2 g_I(\phi_1\phi_2^* +
\phi_1^*\phi_2)\} ( i\bar{\chi}\psi + i\bar{\psi}\chi) \cr
&\ -{1\over 2}\{ g_I[(\phi_1)^2 -(\phi_2)^2] +2 g_R \phi_1\phi_2 \}
(  i\bar{\psi}\psi^* -i\bar{\chi}\chi^* ) \cr
&\ -{1\over 2}\{  g_I[(\phi_1^*)^2-(\phi_2^*)^2] +2g_R\phi_1^*\phi_2^*\}
( i\bar{\psi}^*\psi -i\bar{\chi}^*\chi  ) \cr
&\ -\{  g_R[(\phi_1)^2 -(\phi_2)^2 ] - 2 g_I \phi_1\phi_2 \}
i\bar{\chi}\psi^* \cr
&\ -\{g_R [(\phi_1^*)^2 -(\phi_2^*)^2] -2g_I\phi_1^*\phi_2^* \}
i\bar{\chi}^*\psi      \cr}\eqno\eq $$
After we  eliminate the auxiliary fields,  the $N=2$ supersymmetric
lagrangian (2.12)  becomes  a sum,
$$ {\cal L}_{N=2} = {\cal K} + {\cal L}_{CD} + {\cal L}_\lambda +
{\cal L}_F + {\cal L}_{fi} \eqno\eq $$

For a moment, let us consider the part of the  lagrangian (3.7)
which depends on bosonic fields only. Besides the  kinetic terms, there
are self-interacting terms from
Eqs.(3.3) and (3.4). If $m=v^2 g$ with a real number $v^2$, the last
term in Eq. (3.4) vanishes. If $|g|= e^2/\kappa$, there is a partial
cancellation between the first term of Eq. (3.3) and the third term of
Eq. (3.4), which implies that the phases of $\phi_1$ and $\phi_2$ can
be rotated independently in the bosonic part of the lagrangian (3.7).
Let us restrict ourselves from now on to the case where coupling
constants satisfy
$$ \eqalign{ &\ |g| = {e^2 \over \kappa} \cr
             &\ m = v^2 g \cr} \eqno\eq $$
where $v^2$ is a real number.
The Higgs potential is given by the bosonic part in
the sum of $-{\cal L}_{CD}$ and $-{\cal L}_F$, which is
$$  U(\phi_a) = {e^4\over \kappa^2}|\phi_a|^2 \bigl[ (|\phi_a|^2)^2
+ 2v^2 (|\phi_1|^2-|\phi_2|^2) + v^4 \bigr] \eqno\eq $$
Note that $U(\phi_a) \ge 0 $ regardless of the sign of $v^2$. Without
losing any generality, we will assume that $v^2 > 0$.  The Higgs
potential has two degenerate minima,  the  symmetric one
where $<\phi_a> = 0$ and the asymmetric one  where  $<\phi_1> =0,
<\phi_2> = v$.

Let us now examine the fermionic parts of the lagrangian (3.7).  The
fermion mass term in the symmetric phase is the $m$ dependent terms in Eq.
(3.6). One can see easily that the phases of $\psi,\chi$ can be
rotated independently if $m$ is pure imaginary, which with Eq.(3.8)
implies that $g = \pm ie^2/\kappa$.
Let us choose the positive sign as one can see that  choosing the
negative sign is  equivalent to our choice if $\psi$ and $\chi$ are
exchanged.  With our  choice of the coupling
constant,
$$  g = {m \over v^2} = {ie^2 \over \kappa} \eqno\eq $$
the fermionic part of ${\cal L}_\lambda + {\cal
L}_{CD}+ {\cal L}_{fi}$ becomes the Yukawa interaction,
$$ \eqalign{ {\cal Y}   =  &\ -{e^2 \over \kappa}
v^2 (i\bar{\psi}\psi -i\bar{\chi}\chi) + {3e^2\over \kappa}(|\phi_1|^2
+ |\phi_2|^2)(i\bar{\psi}\psi + i\bar{\chi}\chi) \cr
&\
 -{4ie^2\over \kappa} (\phi_1\bar{\psi} -\phi_2\bar{\chi})(
\phi_1^*\psi -
\phi_2^*\chi) \cr
&\ -{ie^2\over \kappa}(\phi_1\bar{\psi}-\phi_2\bar{\chi})
(\phi_1\psi^* -\phi_2\chi^*)
 -{ie^2\over \kappa}(\phi_1^* \bar{\psi}^*-\phi_2^*\bar{\chi}^*)
(\phi_1^*\psi - \phi_2^*\chi) \cr} \eqno\eq $$

Finally the  $N=2$ supersymmetric
lagrangian (3.7)  becomes the sum  ${\cal L} = {\cal K} -U + {\cal Y}
$, which is given  explicitly by
$$ \eqalign{ {\cal L} =&\ {\kappa\over 2} \epsilon^{\mu\nu\rho}A_\mu
\partial_\nu A_\rho - |D_\mu \phi_a|^2 -{i\over 2
}\bar{\psi}\gamma^\mu \tilde{D}_\mu
\psi -{i\over 2} \bar{\chi}\gamma^\mu \tilde{D}_\mu \chi \cr
&\ - {e^4\over \kappa^2}|\phi_a|^2[ (|\phi_a|^2)^2
+ 2v^2 ( |\phi_1|^2-|\phi_2|^2) + v^4]
-{e^2\over \kappa} v^2 (i\bar{\psi}\psi - i\bar{\chi}\chi)
\cr
&\ + {3e^2 \over \kappa}(|\phi_1|^2 + |\phi_2|^2)
 (i\bar{\psi}\psi +i\bar{\chi}\chi)
 - {4ie^2\over \kappa} (\phi_1\bar{\psi} -\phi_2\bar{\chi})(
\phi_1^*\psi -\phi_2^*\chi) \cr
&\ -{ie^2\over \kappa}(\phi_1\bar{\psi}-\phi_2\bar{\chi})
(\phi_1\psi^* -\phi_2\chi^*)
 -{ie^2\over \kappa}(\phi_1^* \bar{\psi}^*-\phi_2^*\bar{\chi}^*)
(\phi_1^*\psi - \phi_2^*\chi) \cr} \eqno\eq $$
where $\tilde{D}_\mu$ is defined so that $ \bar{\psi}\tilde{D}_\mu
\chi = \bar{\psi} D_\mu \chi - (D_\mu  \bar{\psi} )\chi$ and
${\cal L}$ is hermitian.  Let us consider the internal symmetry of
this lagrangian. There is a $U(1)$ local gauge symmetry inherited from
the four dimensional super QED, which rotates simultaneously
the phases of $\phi_1, \phi_2, \psi,\chi$.  There is a
new global $U(1)$ symmetry due to our special choice of coupling
constants in Eq.(3.10)  which  rotates the phases of
$\phi_1,\phi_2^*,\psi,\chi^*$ uniformly.  The N=2 extended
supersymmetric theory in Ref.[2]  can be obtained easily from the
lagrangian (3.12) by discarding  the terms depending on
 $\phi_1$ and $\chi$.

Let us ask what happens to the supersymmetric transformation of the
lagrangian (3.12). To get the on-shell $N=2$ supersymmetric
transformation, we remove the auxiliary fields from the supersymmetric
transformation by using their equations of motion with the special
choice of the coupling constants. Eqs.(2.6),  (3.1) and (3.10) lead to the
supersymmetric transformation of the gauge field,
$$ \eqalign{ \delta A_\mu = &\ {e\over \kappa} \bar{\alpha}_1
\gamma_\mu ( \psi \phi_2^* -\psi^*\phi_2 -\chi \phi_1^* +\chi^* \phi_1 )
\cr &\  +{e\over \kappa}\bar{\alpha}_2\gamma_\mu(-\psi \phi_1^* +
 \psi^*\phi_1 -\chi \phi_2^*  + \chi^*\phi_2 )\cr } \eqno\eq $$
This supersymmetric transformation is compatible with the local gauge
symmetry.  However the new global transformation which rotates the
phases of $\phi_1, \psi, \phi_2^*, \chi^*$ uniformly is not compatible
with the $\alpha_1$ dependent part of the supersymmetric transformation
because $A_\mu$ is neutral. To remedy this conflict, let us assume
that the parameter $\alpha_1$ is a  complex rather than Majorana spinor
and transforms like $(\phi_1)^2$ under the new global transformation.
Then the supersymmetric transformation for the gauge field can be
modified to be consistent with  the new global symmetry and is given by
$$ \eqalign{ \delta A_\mu = &\ {e\over \kappa}\bar{\alpha}_1
\gamma_\mu( \psi \phi_2^*  + \chi^* \phi_1) - {e\over \kappa}
\bar{\alpha}_1^*\gamma_\mu ( \psi^*\phi_2 + \chi \phi_1^* )
\cr &\  +{e\over \kappa}\bar{\alpha}_2\gamma_\mu(-\psi \phi_1^* +
\psi^* \phi_1 -\chi \phi_2^*  +\chi^* \phi_2  )\cr } \eqno\eq $$
Similarly one can modify the supersymmetric transformations for the
scalar and fermion fields so that they are compatible with the new
symmetry. They become
$$ \eqalign{ &\ \delta \phi_1 = i (\bar{\alpha}_1^*\chi +\bar{\alpha}_2\psi )
\cr
&\  \delta \phi_2 = i (-\bar{\alpha}_1 \psi + \bar{\alpha}_2 \chi )
\cr} \eqno\eq $$
and
$$ \eqalign{  \delta \psi =&\ - \gamma^\mu \alpha_1  D_\mu \phi_2 +
   \gamma^\mu \alpha_2 D_\mu \phi_1
 \cr
&\ +{e^2 \over \kappa} (\alpha_1\phi_2 -\alpha_2\phi_1)( v^2 +
|\phi_1|^2 -|\phi_2|^2) \cr
&\ + {2e^2 \over \kappa} (\alpha_1^* (\phi_1)^2 \phi_2^* +
\alpha_2 \phi_1 |\phi_2|^2)  \cr}  \eqno\eq $$
$$  \eqalign{ \delta \chi = &\  \gamma^\mu \alpha_1^*D_\mu \phi_1
+ \gamma^\mu \alpha_2  D_\mu \phi_2 \cr
&\ + {e^2\over \kappa}( \alpha_1^* \phi_1 + \alpha_2 \phi_2)(v^2 +
|\phi_1|^2 -|\phi_2|^2) \cr
&\ + {2e^2 \over \kappa } ( - \alpha_1 \phi_1^* (\phi_2)^2 +
\alpha_2 |\phi_1|^2 \phi_2 ) \cr} \eqno\eq $$
It is straightforward to  show that the lagrangian (3.12) is
invariant under the supersymmetric transformation given by Eqs.
(3.14), (3.15), (3.16) and (3.17). Since $\alpha_1$ is now a  complex
spinor, the supersymmetry has been extended to $N=3$. The lagrangian (3.12) is
the $N=3$ supersymmetric Chern-Simons-Higgs lagrangian.

\endpage

\chapter{ Elementary Properties }

The energy functional of the bosonic part is given by
$$ E = \int d^2x \bigl[ |D_0 \phi_a|^2 + |D_i\phi_a|^2 + U \bigr] \eqno\eq $$
With the Gauss law for the bosonic part,
$$ \kappa F_{xy} = -i(D_0\phi_a^\dagger\phi_a - \phi_a^\dagger D_0\phi_a)
\eqno\eq $$
one can write the energy functional as
$$ \eqalign{ E = \int d^2x &\ \biggl[ |D_x\phi_a \pm i D_y \phi_a|^2 +
|D_0\phi_a  \pm  {i e^2 \over \kappa} \phi_a (|\phi_a|^2 -v^2)|^2 \biggr.  \cr
&\ + \biggl.
 {4e^4\over \kappa^2} v^2|\phi_1|^2|\phi_a|^2  \bigr]  \mp ev^2 \Psi \cr}
\eqno\eq $$
where $\Psi = \int d^2 x F_{xy} $ is the total magnetic flux.
Eq.(4.3) implies a  bound on the energy functional,
$$ <E> \,\, \ge \,\, ev^2 |\Psi| \eqno\eq $$
The energy  bound is saturated by configurations satisfying
the Gauss law and the following equations,
$$ \eqalign{ &\ \phi_1 = 0  \cr
 &\ D_1\phi_2 \pm i D_2 \phi_2 = 0 \cr
 &\ D_0 \phi_2 \pm {ie^2 \over \kappa} \phi_2 (|\phi_2|^2-v^2) = 0 \cr
} \eqno\eq $$
The equations for $\phi_2$ are identical to those of the self-dual
Chern-Simons system. In this theory,  there are nontopological solitons in
symmetric phase and topological vortices in asymmetric phase
saturating the energy bound as shown in Ref. [1].

What are the elementary excitations of the theory (3.12)?
In the symmetric phase,  the quadratic terms in (3.12) with gauge
interaction become
$$ \eqalign{ {\cal L}_{symm} = &\  {\kappa \over 2} \epsilon^{\mu\nu\rho} A_\mu
\partial_\nu  A_\rho - |D_\mu \phi_a |^2 -i \bar{\psi} \gamma^\mu
D_\mu \psi - i\bar{\chi}\gamma^\mu D_\mu \chi  \cr
&\
- {m^2 \over 2}  |\phi_a|^2 - { m \over 2} (i\bar{\psi}\psi -
i\bar{\chi} \chi ) \cr} \eqno\eq $$
where $m \equiv 2 e^2v^2/ \kappa$. There is no particle excitation
related to the gauge field. There are two complex scalar and two spinor
fields of mass $m/\sqrt{2}$. It is well known that charged particles
carry fractional spin and satisfy fractional statistics due to the
Chern-Simons term. The excitation related to the scalar field has spin
$ s= 1/(4\pi k)$.  The sign difference between the two mass terms of
$\psi$ and $\chi$ implies that the spins of $\psi$ and $\chi$ are $ s+
1/2$ and $s -1/2$, respectively.

In the asymmetric phase, we use the unitary gauge where $\phi_2 =
f/\sqrt{2} + v$ with a real $f$. By neglecting matter self-interaction, we get
from
Eq.(3.12)  a lagrangian
$$ \eqalign{ {\cal L}_{asym} =   &\
{\kappa \over 2} \epsilon^{\mu\nu\rho} A_\mu
\partial_\nu  A_\rho -  |\partial_\mu \phi_1 |^2 - {1\over 2} (\partial_\mu
f)^2
 -i \bar{\psi} \gamma^\mu \partial_\mu \psi \cr
&\  - {i \over 2} [ \bar{\chi}_R \gamma^\mu
\partial_\mu \chi_R + \bar{\chi}_I\gamma^\mu \partial_\mu \chi_I ]
-{\kappa m \over 2} (A_\mu)^2 - m^2 |\phi_1|^2 - {m^2 \over 2} f^2 \cr
&\  +m i\bar{\psi}\psi - {m\over 2} (
i\bar{\chi}_R\chi_R - i\bar{\chi}_I\chi_I) \cr
&\ + e A_\mu [ -i(\partial^\mu \phi_1^* \phi_1 - \phi_1^* \partial^\mu
\phi_1) + \bar{\psi}\gamma^\mu \psi + i\bar{\chi}_R\gamma^\mu \chi_I )\cr &\ -
e^2 (A_\mu)^2 [ |\phi_1|^2 + {1\over 2} ( 2\sqrt{2}e^2vf +
e^2f^2) ]
 \cr}
\eqno\eq $$
where $\chi = \chi_R + i\chi_I$ with $\chi_{R,I}$ being Majorana
spinors.  Due to the Higgs mechanism, there is a physical degree of
freedom for $A_\mu$, with  mass $m$ and spin $1$.  There are also three
spin $1/2$ particles for $\psi, \chi_I$, three spin zero particles for
$f, \phi_2$, and one spin $-1/2$ particles for $\chi_R$, all with the
same mass $m$.

\endpage

\chapter{Symmetry Algebra}

Let us investigate the symmetries of the lagrangian (3.12). The local
gauge symmetry implies  the Gauss law constraint,
$$ {\cal G} \equiv \kappa F_{xy} +   e\rho_c =0, \eqno\eq $$
where the electric charge density is given by
$$ \rho_c = i  (D_0\phi_a^* \phi_a - \phi_a^*D_0\phi_a)
-\psi^\dagger\psi -\chi^\dagger \chi .\eqno\eq $$
The local gauge transformation is generated by the operator ${\cal G}
$.  By integrating Eq.(5.1) over space, we see that the total electric
charge $Q = \int d^2x \rho_c$ and total magnetic flux $\Psi = \int
d^2x F_{xy} $ are related by
$$\kappa \Psi = - eQ .\eqno\eq $$
There is an additional global symmetry mentioned in section 3. The
total charge for this symmetry, which transforms the phases of
$\phi_1, \psi, \phi_2^*, \chi^*$ uniformly,  is given by
$$   T = \int d^2x \biggr[  i(D_0\phi_1^* \phi_1 -
\phi_1^*D_0\phi_1)  - i(D_0\phi_2^* \phi_2 - \phi_2^* D_0\phi_2 )
 -\psi^\dagger \psi + \chi^\dagger\chi \biggr] . \eqno\eq
$$

There are also spacetime symmetries. The symmetric energy momentum
tensor is given by
$$\eqalign{  T_{\mu\nu} =&\  D_\mu \phi_a^* D_\nu \phi_a+ D_\nu \phi_a^* D_\mu
\phi_a
  +{i \over 4} \bar{\psi}(\gamma_\mu \tilde{D}_\mu + \gamma_\nu
\tilde{D}_\nu )\psi \cr
&\  + {i\over 4} \bar{\chi}(\gamma_\mu \tilde{D}_\nu
+ \gamma_\nu \tilde{D}_\mu) \chi + \eta_{\mu\nu}{\cal L}
\cr } \eqno\eq
$$
The generators for the spacetime translation  are  the  conserved three
momentum,
$$ {\cal P}^\mu = \int d^2x T^{\mu 0} \eqno\eq $$
The generators of the Lorentz transformation form a three vector,
$$ {\cal M}_\mu = \int d^2x \epsilon_{\mu\nu\rho} x^\nu T^{\rho 0}
\eqno\eq$$
In particular, the total angular momentum is  given by
$$ J = {\cal M}^0 = \int d^2x (x^1 T^{20} - x^2 T^{10} ) \eqno\eq$$

There are three independent generators for supersymmetric
transformations, given by $\delta \phi_a = i[ i\bar{\alpha}_1{\cal R}
+ i\bar{{\cal R}}\alpha_1 +\bar{\alpha}_2 {\cal S}, \phi_a] $, etc.
One can obtain these generators by using the supersymmetric transformations
(3.14), (3.15), (3.15) and (3.16).  Corresponding to the complex
spinor $\alpha_1$, there is a complex spinor operator,
$$ \eqalign{ {\cal R} = \int d^2 x &\ \biggl\{   \bigl[D_\mu \phi_2^*
\gamma^\mu + {e^2 \over \kappa} \phi_2^*  (v^2 + |\phi_1|^2 - |\phi_2|^2 )
\bigr]\gamma^0 \psi + {2 e^2 \over \kappa} (\phi_1)^2 \phi_2^*
\gamma^0 \psi^*
\biggr.
\cr
&\ -  \biggl. \bigl[ D_\mu \phi_1 \gamma^\mu - {e^2 \over \kappa}\phi_1 (v^2 +
|\phi_1|^2 - |\phi_2|^2)
\bigr]\gamma^0\chi^*  - {2e^2 \over \kappa} \phi_1 (\phi_2^*)^2
\gamma^0 \chi  \biggr\} \cr }
\eqno\eq $$
For the Majorana spinor parameter $\alpha_2$, there is a Majorana generator
$$ \eqalign{ {\cal S} = \int d^2x &\ \biggl\{ -\bigl[ D_\mu \phi_1^* \gamma^\mu
+ {e^2 \over \kappa}\phi_1^*(v^2 + |\phi_1|^2 -3 |\phi_2|^2)
\bigr]\gamma^0 \psi  \biggr.  \cr
&\ \biggl.  -\bigl[ D_\mu \phi_2^* \gamma^\mu -{e^2\over \kappa}
\phi_2^* (v^2+ 3|\phi_1|^2 - |\phi_2|^2)
\bigr] \gamma^0\chi  + h.c. \biggr\}
 \cr} \eqno\eq $$

Let us consider the algebra of symmetry operators. The equal time
commutation relation between fields are standard, for example, $[
A_1(\vec{x}) , A_2(\vec{y})] = (1/\kappa) i \delta (\vec{x} - \vec{y})
$. The interesting and nontrivial ones  are those  between supersymmetric
operators. The complex spinor generator ${\cal R}$ satisfies,
$$ [i\bar{\alpha}_1{\cal R}, i\bar{{\cal R}}\beta_1] =
\bar{\alpha}_1\gamma^\mu \beta_1 {\cal P}_\mu +  i\bar{\alpha}_1\beta_1 Z
\eqno\eq $$
where the central charge $Z$ is given by
$$    Z = -\int d^2x \bigl[ eF_{xy}( |\phi_1|^2  -|\phi_2|^2)
 + {e^2 \over \kappa} \rho_c
(v^2 + |\phi_1|^2 - |\phi_2|^2)  \bigr]  \eqno\eq $$
The Gauss law constraint (5.1) implies that
the  central charge can be expressed as
$$ Z =  ev^2 \Psi. \eqno\eq $$
The Majorana  supersymmetry operator ${\cal S}$ satisfies
$$ [ i\bar{\alpha}_2{\cal S} , i\bar{\cal S} \beta_2] =
2\bar{\alpha}_2\gamma^\mu \beta_2 {\cal P}_\mu .\eqno\eq $$
Under the global transformation $T$,  only ${\cal R}$ transforms
nontrivially,
$$ [ T, {\cal R} ] = 2 {\cal R}. \eqno\eq $$
As said  in Sec.3, the operator ${\cal R}$ transforms like
$(\phi_1)^2 $ under the new global transformation. Thus, the  above
symmetry operators form an $N=3$ extended superalgebra.

The physical implication of the central charge $Z$ can be reached by
considering
$$ \eqalign{ {\cal R}_\pm  = {1 \pm i\gamma^0 \over 2} &\  {\cal R}\cr
 =   \int d^2x \biggl\{ &\ {1 \pm i \gamma^0 \over 2}[ - \psi(
D_0\phi_2^* \pm i
 \phi_2^* X )  + \chi^* (D_0\phi_1 \mp i  \phi_1 X ) ] \biggr. \cr
 &\ \pm i { \gamma^1 \pm i\gamma^2 \over 2}[ \psi(D_1 \phi_2^* \mp i
D_2 \phi_2^*) - \chi^* (D_1\phi_1 \mp i D_2 \phi_1 ) ] \cr
&\ \biggl. \pm i {1 \pm i \gamma^0 \over 2} [ - \psi^* {2e^2\over
\kappa}(\phi_1 )^2\phi_2^* +
i \chi {2e^2 \over \kappa}
\phi_1(\phi_2^*)^2 ]\biggr\} \cr }\eqno\eq $$
where $X = v^2 + |\phi_1|^2 - |\phi_2|^2 $.
After multiplying $1 \pm i\gamma^0 /2 $ to Eq. (5.11) and taking trace,
we obtain
$$ {\cal P}^0 = \mp Z + \sum_A \{ {\cal R}_{\pm A}, {\cal R}_{\pm A}^\dagger
\}
\eqno\eq $$
Hence there is a bound on the expectation value of the energy,
$$ <{\cal P}^0> \,\, \ge  \,\,\, | < Z > |  \eqno\eq $$
The inequality is saturated if the coefficient of each fermion field
in Eq.(5.16) vanishes. These conditions are independent of the
sign of $v^2$ and can be easily shown to be equivalent to the
self-dual equations (4.4) for positive $v^2$. (For  negative $v^2$,
one has to rederive the self-dual equations. )

What are the possible representations under this $N=3$ supersymmetric
algebra?  Let us consider only massive states. In the rest frame where
$P^\mu = (M,0,0)$,  Eqs.( 5.11) and (5.14) become
$$ \eqalign{ &\ \{ {\cal R}_{\pm 1}, {\cal R}_{\pm 1}^\dagger \} = M \pm Z
\cr
&\  \{ {\cal S}_1, {\cal S}_2 \} = 2M  \cr } \eqno\eq $$
Note that ${\cal R}_{\pm 2} = \pm i {\cal R}_{\pm 1}$ from Eq. (5.16).
After a suitable normalization, we can build fermionic creation and
annihilation operators.  By applying them on a Clifford vacuum, one
can construct a representation.
Let us first consider charged states which saturate the
bound (5.18).  From a Clifford vacuum $|\Omega>$ of angular momentum
$j$, one can build two $j+1/2$ states and one $j+1$ state. Elementary
particles in the symmetric phase saturate the energy bound and form
such a representation.  For nontopological solitons in symmetric phase
and topological vortices in asymmetric phase, the classical solution
saturates the energy bound. Hence, one expects these solitons to form
the same representation as do  elementary particles in the
symmetric phase. This remains to be seen since the quantization of
solitons are somewhat involved.  Let us now consider states which
do not saturate such bound.  An irreducible representation for those
states is made of
one angular momentum $j $ state, three  $j+ 1/2$ states, three
 $j+1 $ states, and one $j+3/2$ states.  Elementary
excitations in asymmetric phase form such a representation.

With the  Chern-Simons term as the only gauge kinetic term, there
is no physical degree for the gauge field in the symmetric phase. In
the broken phase there is only one massive vector boson of spin $1$
and it is neutral. A neutral nontrivial representation of the $N=4$
supersymmetry is made of one spin 1, four spin 1/2, six spin 0, four
spin -1/2, and one spin -1 states. As there is no candidate for spin
$-1$ state, there is no $N=4$ supersymmetric Chern-Simons Higgs theory
with a single gauge field. Hence our $N=3$ theory is the `maximally
supersymmetric' theory with a single gauge field and no gravity.

\endpage

\chapter{ Conclusion}

We have constructed an $N=3$ extended supersymmetric
Chern-Simons-Higgs theory and studied its properties. In addition to
the original abelian gauge symmetry there exists a new global $U(1)$
symmetry. The self-dual Chern-Simons Higgs systems studied
before\refmark{\rJW, \rLLW} are part of this $N=3$ supersymmetric
theory. This theory turns out to be the maximally supersymmetric with
a single gauge field and no gravity.

Let us conclude by mentioning several open questions.  The $N=4$
supersymmetric nonabelian theory in four dimensional spacetime is the
maximal supersymmetric renormalizable theory and is  known to be free of
ultraviolet divergence.  As the $N=3$ theory discussed in this paper
is maximally supersymmetric in three dimensional spacetime, it would
be intersting to find out whether this theory is also ultraviolet
finite.

One very interesting question is about the finite shift of the
coefficient of Chern-Simons term, which is supposed to be due to
fermion one loop correction\refmark{\rColeman}. In the symmetric phase, the two
fermions have opposite mass and there is no correction. In the
asymmetric phase one of the fermion mass terms changes as shown in
Sec.4, leading to non zero shift of the coefficient.
 It would be
interesting to find out whether there is nontrivial contribution from
Higgs field to keep the total contribution to be identical to that in
the symmetric phase.

\centerline{\bf Acknowledgement}
We thank V.P. Nair for stimulating discussions.

\endpage
\refout

\end